
\hfill{FERMILAB-PUB-93/15-T}\smallskip
\hfill{March 1993}\smallskip
\hfill{Revised: January 1994}\smallskip
\vskip 0.5in
{\bf The Thermodynamics and Economics of Waste}
\bigskip\bigskip
Dallas C. Kennedy, Research Associate,\medskip
Fermi National Accelerator Laboratory,\medskip
P.O. Box 500 MS106, Batavia, Illinois 60510 USA$^*$
\vfootnote*{Present address: Department of Physics, University of Florida,
Gainesville, Florida 32611 USA}
\bigskip\bigskip\bigskip
\centerline{\bf Abstract}
\bigskip
\noindent The increasingly relevant problem of natural resource use and
waste production, disposal, and reuse is examined from several viewpoints:
economic, technical, and thermodynamic.  Alternative economies are studied,
with emphasis on recycling of waste to close the natural resource cycle.  The
physical nature of human economies and constraints on recycling and energy
efficiency are stated in terms of entropy and complexity.
\bigskip\bigskip
{\it What is a cynic?  A man who knows the price of everything, and the value
of nothing.}\medskip
\hfill{\it Oscar Wilde}~[1]
\bigskip\bigskip
Our planet is finite in size and, except for a few energy and matter flows,
its biosphere forms a closed system.  The envelope that makes
life possible extends a small distance below the surface of the Earth and
less than a hundred miles into the atmosphere.  While almost closed, the
biosphere is not static, but is constantly changing, moving flows of energy,
air, water, soil, and life around in a shifting, never-repeating pattern.
The combination of general physical laws and the specific properties of the
Earth places important constraints on the activities of life, some embodied
in the metabolism and forms of living creatures, others imprinted into their
genes by selective effects.  All life needs sources
of energy for sustenance and imposes a burden of waste on its environment.
Since this waste is usually harmful to the creatures emitting it, the
environment must, if these creatures are to continue living, break the waste
down into less toxic forms and possibly reuse it.

The growing dominance of humankind over the planet, both by technological
power and by numbers, imposes certain costs on the biosphere, sometimes
a result of
conscious attempts at controlling Nature, but more frequently by unwitting
influence.  Moreover, the burden of carrying the
activities of human sustenance, unlike that of other animals, cannot be
understood by considering the physical and biological activities of each
person in isolation.  Because of their
unique position at the top of the food chain, their tool-making abilities,
and the co-operative character of human activities (the division of labor or
specialization),
the physical and biological aspects must be considered together in the context
of the peculiarities of economic life~[2].  The economic aspects take on
an independent importance because of the absence of any automatic, given means
of human subsistence and the extension of individual self-sufficiency
by surplus production and trading~[3].  This point of
view is necessary for comprehending the ecological significance of all
economies
more elaborate than the simplest subsistence or household
economies, up to and
including the most sophisticated systems of technology and trade.  On the
other hand, the formulation of economic theory has generally taken place, apart
from a few outstanding exceptions~[4], with scant
attention paid to how physical and biological laws make their appearance in the
course of humanity's reshaping of its environment~[5].  This article
explores human economics as a special branch of ecology, with particular regard
to the production, disposal, and possible reuse of waste.
\bigskip
\noindent{\bf Natural Resources and Rents}
\bigskip
To begin investigating the problem that the waste produced by our economic
activity poses, we consider, in simplified diagrammatic form, the streams
of economic activity that make up a conventional economy (Figure~1)~[6].
We ignore complications such as hoarding, credit and debt, and
unemployment of resources, as they do not change the essentials.
One set of streams is made up by various flows of matter and energy in the form
of goods and services, a collection
usually called the {\it real} economy.  The inputs of the real economy are
natural resources, labor, and capital (tools)~[7].
The outputs are consumption goods,
capital goods, and waste.  Two of the output streams are tied back to the
input streams to form a closed system, the consumption goods used by labor
to sustain and reproduce life and the capital goods added to the previously
existing stock of capital.  The third output stream is waste, discharged
into the environment, but not tied back to the input stream of natural
resources, which are taken from the environment.  Since a functioning economy
is never exactly in an equilibrium of steady flows, but always expanding or
contracting, diagrams such as Figure~1 are only instantaneous snapshots.
At any moment, the resources available (human and otherwise) are limited, but
this limit generally rises and falls over time.

Counter to the flow of matter and energy, in an exchange system, is the
{\it money} or {\it information} economy of income streams.  The income flow
corresponding
to a particular item is the product of the physical flow of an item
and its price: prices are information~[8].  To the inputs are
paid three types of income: wages to labor, profit to capital, and
{\it rents} to the owners of natural resources~[3].  The term ``rent'' should
not be confused with its common usage; in technical economic usage,
``rent'' refers to the monopoly incomes received by the owners of scarce and
non-substitutable natural resources --- such as land, in the common meaning.
Two of these incomes go to pay for two of the
outputs, wages to pay for consumption goods (consumption spending), profits
to pay for capital goods (investment spending).  The rent incomes are
simply {\it gratis} and pay for nothing.  In fact, rents are extracted from
the flow of wages and profits as a tax and necessarily diminish the
level of consumption and investment, as first pointed out by David
Ricardo~[9,10].

The scarcity of land and of the food produced on it led to the first
crisis of the infant industrial system in England in the early
nineteenth century, a crisis that played a prominent role in
the works of Ricardo and
his fellow economist, Thomas Malthus~[9,10,11,12].
As the industrial system
and the population of workers expanded, it ran into the fixed amount
of land in England and the
relatively fixed level of food production possible on that land.  The rent
on scarce land threatened to destroy the profits on capital and halt
investment; rising food prices apparently condemned the workers to a
chronically low level of subsistence.  The inability of England to feed
itself beyond a certain limit formed the centerpiece of a powerful argument
for free trade in agriculture; what food England could not produce, it would
import and pay for with manufactured goods.  Eventually, the arguments of
Ricardo and others won out, and England adopted free trade in food in 1846.
Other factors also prevailed to prevent a fatal crisis.  Population growth in
England was slowed by delayed marriages, urbanization,
and, later, by the introduction of
artificial birth control methods.
Improved agricultural methods made much
higher levels of production possible on a given amount of land.  A larger
population had the advantage of greater diversity, specialization, and
productivity, given enough freedom to make economic improvements.
A final factor
of great importance in preventing a crisis of resource exhaustion has been the
substitution of more common inputs for scarcer ones~[13].  Few famines,
at least in recorded history, seem to be due to overpopulation,
but rather are caused by warfare, panic hoarding, and criminal or misguided
governments.
However, it is misleading to concentrate only on food production, as
overpopulation in localized ecosystems
is responsible for other environmental problems,
such as deforestation, soil erosion, and drastic changes in rain cycles.

Later in the nineteenth century, the American amateur economist and
crusader Henry George rediscovered the importance of rents on land and
launched an attack on them as the source of society's economic ills in his
once-famous work, {\it Progress and Poverty}~[10,14].  Although George
exaggerated the evil of rents, he performed a public service by
raising the issue of unearned rent incomes and by emphasizing the difference
between unearned monopoly rents on fixed land
and profit on competitive capital, which the
common use of the term ``profit'' confuses.
His cure for the unearned incomes derived from monopoly control of fixed
natural resources was to introduce a single tax that would absorb all rents.

Although the monopoly rent incomes received by natural resource owners seem
like
an unjust burden on the rest of society, rents cannot be condemned out of hand
as serving no function.  In particular, in an ecologically conscious age, we
should recognize rents and private ownership of resources
as brakes on the overexploitation of Nature.  If people
have to pay for resources, they will use less of them.
This is all the more true if the resource owners impose extra charges,
{\it reservation prices,} to cover not only the current cost of
resource extraction but also to prevent future depletion~[6].  The charging
of interest on credit (future discounting) has the same effect.  Many of
the most prominent examples of the overuse of resources today --- overlogging
in the Pacific Northwest, overgrazing on the Great Plains, wasteful water
usage in California, suburban overdevelopment, overuse of chemicals in
agriculture, urban overconsumption of
food and discouragement of agriculture in the Third World,
destruction of the topical rainforests, environmental devastation
in the Communist countries --- are supported by
public subsidies designed to keep raw materials cheap to their users~[15].
The result is the {\it tragedy of the commons:} if everyone is
supposed to take care of something, no one takes care of it~[16].  Furthermore,
by shielding the users of a particular resource from the full cost of their
exploitation, these subsidies not only encourage present overuse of resources,
but also discourage the search for cheaper (less scarce) substitutes and
thereby exacerbate future shortages.
Although environmentalists are not usually thought of as friends of private
property, private ownership of a resource provides a powerful motive to
bar others from its overuse and is usually the best way to protect wilderness
areas~[17].  Private property allows, while limiting, the
individual exploitation of Nature by separating ownership from political
power.  Of course, private ownership is not a panacea;
some owners are as shortsighted as anyone and can strip their property
for quick benefit.
Nonetheless, if societies are interested in protecting
the environment, they could do worse than refrain from these
subsidies.  Such discriminatory intervention subsidizes select
groups of raw materials users at the expense of everyone else and of the
environment.  Natural resource use is always subject to
diminishing returns and thus has no economic justification for subsidy, unlike
knowledge- or technology-based sectors of an economy that sometimes exhibit
increasing returns to scale spread out over time~[18].
Unfortunately, these subsidies are usually of great benefit to select,
politically powerful groups whose influence over
governments is difficult to remove.  The subsidies result from the
the conversion of the profit motive, the basis of any functioning
economic system, into the power motive characteristic of politics~[19].

In a conventional economy, no money exchange is associated with the other
end of the resources flow, the production of waste.  The waste flow is
discharged into the environment with no cost or payment.  In short, the price
of waste is zero.
\bigskip
\noindent{\bf The Pay-for-Trash-Removal Economy}
\bigskip
The conventional economy is not the only possible one:
waste in many cases is not dumped for free into
the environment.  Instead, the waste producers pay someone else to remove
their trash and to dispose of it in some fashion (in a landfill, for example)
that often involves an additional charge.  This type of economy is
diagrammed in Figure~2.  Money payments accompany the waste outward to pay
for its disposal.

But the trash-removal economy is odd, since the waste producers pay
{\it twice} for their use of natural resources, once for the inputs and
once again for waste removal.  So there are now {\it two} rents extracted from
the economy, both rents being subtracted from the flow of wages and profits.
Such a system will eventually press up against a new Ricardian barrier,
because the {\it producers} of waste,
unlike producers of consumption and capital goods, pay the {\it consumers}
of waste for its removal, rather than the other way around.
Waste has a negative price in this type of economy.

Of course, from a more limited point of view, this arrangement makes sense
to the parties involved.  The waste removers receive a rent income, while
the waste producers have their trash removed.
But as the level of waste production rises with
economic growth, and the free space for waste disposal declines, the rent
charges for waste removal must necessarily rise and put a larger and larger
burden on the rest of the economy, wages and profits.  If the natural
resource inputs also become scarce, then the economy is caught in an even
worse squeeze, with rent being extracted from both ends.
The desire of select producers for cheap input
resources and enlarged export markets
has already inspired one form of imperialism at various periods in economic
history~[3,20]; the frequent absence of a price system for waste is
now prompting a new ``garbage imperialism'', the search for cheap places
to dump trash~[15].

Unfortunately, the pay-for-trash-removal transaction has become the paradigm
for many other kinds of pollution control.  For example, the control of air
pollution is accomplished typically by direct regulation of pollution sources
with limits on emission.  The cost for the ``trash removal'' is necessarily
borne by wages and profits, either directly or indirectly, in a
clumsy way.  A simpler solution, where the polluted resource is privately
owned, is for its owner to bear the cost, or its polluter to recompense the
owner.  For a polluted commons, such as the air or water, that cannot be
privatized, the {\it legal right} to pollute it can be, such as in a market for
pollution shares, where polluters can trade rights to emission with an overall
fixed number of pollution shares.  But all of these schemes
share the same feature: paying for trash removal.

There is a broad justification for the pay-for-removal approach found in
the economics literature on pollution which rests on the proposition that
by paying for trash removal, a society is ``buying'' a clean
environment~[6,21].
People are behaving ``as if'' there is a market for ``cleanness'', with
so much clean environment available at different costs and varying levels
of social preference for cleanness at different prices.  The difficulty with
this justification is that the pollution of the environment is a product of
human activity in the first place; the environment is not a manufactured
commodity, and Nature is not subject to any but superficial and temporary
human control, nor does it exist only for our use, nor can it in general be
priced by us (except for those parts that we do use and trade with one
another),
because we did not make it.
Furthermore, waste disposal is not subject to substitution the way inputs are,
except for the possibility of finding less wasteful production techniques
that do not reduce useful output.
\bigskip
\noindent{\bf The Recycling Economy}
\bigskip
The way out of this conundrum is to make the natural
resources cycle closed like the two other economic cycles (Figure~3). A
partial solution is to use the rent income from natural resources, which
pays for nothing now, to pay for the trash removal.  Since there would be no
direct physical flow from outputs (trash) to inputs (natural resources),
there would be no market to guide the necessary counterflow of money, and
the payment for trash removal would have to be implemented by a tax on
natural resource income.  Such a tax would be an updated ecological version
of Henry George's universal land tax, instead levied
on the resources that contribute to the production of waste.

The full solution, however, is to start the physical flow of trash output
back to natural resource input --- that is, to recycle.  There would then be
a market for trash, and the counterflow of money to pay for it, derived from
natural resource rents, would be automatic.  Waste output would be treated
like the other outputs, consumption and capital goods, in that the
{\it consumers} of waste (the recyclers) would pay the {\it producers} of
waste for the trash.  The availability of recycled materials for input would
subtract from the income of virgin natural resource owners, and the natural
resource rents would in effect pay for the recycling, establishing full
closure (both physical and monetary) of the resource cycle.

The economics of waste can be restated in
Marshall's graphical formulation~[22], with supply
and demand curves for waste, as in Figure~4.  The supply curve is the
marginal cost of each additional unit of waste as a function of waste flow
(the partial derivative of the cost function with respect to waste flow).
Typically, the marginal cost for waste production is {\it negative,} at least
to start with; that is, the producers of waste find their costs {\it
decreasing}
as they produce more waste associated with the making of their useful outputs.
Eventually, with high enough waste output, the production process
becomes increasingly inefficient, the cost function starts to {\it rise} as a
function of waste flow, and the marginal cost of waste becomes positive.
We assume, as is always the case with natural resource costs,
that the waste supply curve is
subject to diminishing returns and therefore always rises with quantity.  In
the absence of a market for waste (zero utility of waste),
the producers generate waste up to the point where the
marginal cost of waste is zero.  This fact explains
why, even when unregulated, producers in a market system do not produce an
infinite amount of waste.  Such behavior serves as an illuminating
counterexample to the usual pattern in command economies, which, lacking a
price system, generally have no sensible way to account for costs.  Natural
resource exploitation consequently is often subsidized to extreme levels, and
the result is an almost infinite
production of waste and pollution, as reported in great detail by a number
of visitors to Eastern Europe and the former Soviet Union~[23,24].

The demand curve for waste is the marginal utility to the consumers of waste;
that is, the partial derivative of the consumers' utility (reflected by what
they are willing pay for the waste) with respect to waste flow.  As shown in
Figure~4, the marginal utility might be positive for small waste flows --- in
that case, consumers can find some use for the waste --- but the curve usually
becomes negative for large enough waste flows; that is, the curve
is really a marginal {\it dis}utility for waste.
Or, if the waste has no use at all, the demand curve is
always negative (not shown; in that case, the points E and F coincide).
Figure~4 shows the typical result in this
situation.  The market equilibrium for waste --- point of maximum benefit
(utility less cost) to
society --- occurs at {\it negative} marginal
cost and utility, so that waste has a {\it negative} price.
Thus, Figure~4 displays in graphical form our earlier example of the
pay-for-trash-removal economy.  Such a result might be embodied in real
markets, as it is with trash collection services and landfills, or implemented
by direct regulation, as when the government makes known through legislation
or administrative ruling the putative social disutility of waste.  The question
of which approach to use
reduces to the question of creating, wherever possible, markets for the waste.

Careful examination of the market equilibrium reveals the mutually beneficial
nature of the transaction to both producers and consumers of waste, even with
negative prices. Referring to Figure~4,
the area FDCB is what producers pay consumers to take the
waste.  On the other hand, the producers save the area FDCA in costs by
producing the waste; it is clear that this cost saved is
greater than or equal to
what the producers pay for waste removal.  Similarly for the
consumers of waste, the income they receive for removing
the waste is greater than or equal to their
total disutility for the waste, the area EDC, and they may also receive
positive utility from the waste, the area EFG.

The long-term difficulties with waste production discussed earlier can be
restated in this graphical language.  As ecological niches for waste
run out and a society's tolerance for rising cumulative
waste declines, its utility for waste
becomes increasingly negative.  That is, the demand
for waste moves down and to the left in the figure.  This change forces
producers to produce less waste, which can have one of two impacts, or a
combination of both: Either the producers produce less waste by also
producing less useful output, reducing the society's standard of living and
capital accumulation;
or the producers continue to produce the same useful output as before
with less waste by adopting less wasteful techniques.

The recycling economy is
illustrated in Figure~4 by the alternative demand curve for waste, which is
always positive, or at least positive for the waste flows relevant
to this discussion.  That is, in a recycling economy, the waste is useful
and its marginal utility positive.  In this case, the price of waste is
{\it positive.}  As explained
earlier, the income flow to pay for this transaction is ultimately extracted
from the rent incomes of virgin natural resource owners.  Curiously, the
amount of ``waste'' produced in the recycling economy is higher than it is in
the pay-for-trash-removal economy, as Figure~4 shows, simply because the
``waste'' is more useful and so there is a demand for it.  The waste
transaction
in this case is also mutually beneficial to both parties; the graphical
proof follows the same steps as in the case of negative prices above.
\bigskip
\noindent{\bf Limits to Recycling}
\bigskip
The recycling economy already
exists in embryonic form~[15].  Some recycling takes place
of metals, paper, glass and oil.  Certain automobile manufacturers are
planning the reusable car, to end the abandonment of autos in
junkyards.  Junkyards already pay for old cars and
cannibalize them for spare parts.
But the full impact of natural resource scarcity and limits
to waste removal has not yet been felt in advanced economies.  The logical
ideal for the development of a ``green'' (ecologically correct)
economy is: as much recycling as
possible and zero production of non-recyclable wastes.
The ``green'' economy must move as close as possible
to zero waste production without reducing useful output and to the
complete recycling of what waste {\it is} produced.  The key to
recycling is obviously finding ways to make waste useful.

The immediate limits to recycling are institutional.
Almost all producers have set up their production
processes so as to use virgin natural resources only.  They are not organized
to recycle internally their own waste or recycle other
producers' waste.  The ideal of complete recycling would be implemented,
either by each production facility forming a closed system or by many
facilities forming a symbiotic
``food chain'' of producers and recyclers of waste.  Internal
recycling of waste already exists for some extraordinarily valuable inputs,
such as catalysts in chemical reactions that can be used an indefinite number
of times.  An example of external recycling by a consumer-producer loop
is the venerable milk bottle~[25].  A large potential recycling loop exists
with
aluminum cans, since the trucks that deliver the filled cans usually drive
back to the producer without any used cans.
Recycling ``food chains'' are less common.
Cogeneration of heat by burning trash is a primitive example,
although the burning itself pollutes.
The change necessary to implement such schemes mostly involves a change of
mentality, a more accurate recognition of the efficiencies to be gained by
internal recycling and recycling chains.  For the latter, markets would have
to be created to trade the waste.

The technological limits to recycling form a more basic obstacle.  The
limits have two interrelated components: design of original products to
facilitate their later recycling, and development of efficient recycling
techniques~[15].  The ``green'' design of original products requires the
foresight and incentive to anticipate recyclers' needs and limitations
and to incorporate these into the product from the start.  The establishment
of stable and mutually beneficial
relationships between producers and recyclers (or better, producers
becoming their own recyclers) would make this process much easier.
Some soft drink companies already buy back
their own aluminum cans after use by consumers to be recycled for iterated
use.  A well-known example of a recycling bottleneck created by poor
design is the glossy finish used on some types of newsprint that prevents
their reprocessing.  Efficient recycling
techniques have already been developed for aluminum, steel, glass, and some
types of paper and plastic.  Many more materials could be recycled.

Even after institutional and technical obstacles are overcome, however, there
is an ultimate physical limit to recycling, the limit posed by the second law
of thermodynamics, which requires the total entropy of the universe never to
decrease~[26].  Entropy admits of precise definition, either in statistical,
microscopic form or thermodynamic, macroscopic form, but in either case, is
a measure of disorder.  A process of maximal thermodynamic efficiency is
{\it adiabatic} or {\it reversible,} with no net entropy production.  A
less-than-ideal process is irreversible and involves a net increase in
disorder,
which typically appears as waste heat unusable for work; the increment of
entropy is the increment of waste heat divided by the absolute temperature (in
degrees Kelvin) at which is produced.  However, although total entropy never
decreases, it can be moved around in space and time so as to create regions
and periods of greater order, but only by creating at least an equal
amount of {\it dis}order in other places and times.  Useful goods and
activities
are fashioned from raw materials into a more organized form, requiring the
expulsion of entropy in time and/or space that reappears as waste heat and
junk.  To make the growth of organization~[27] more precise, we distinguish two
kinds of entropy, the {\it maximal} entropy or disorder a given macroscopic
object or process {\it could} have, depending only on its size or duration,
parts, and energy; and the {\it actual} entropy or disorder the object or
process {\it actually} possesses.  The former is fixed for a given system,
while the second law requires
the latter never to decrease.  The generalized {\it complexity} or {\it
thermodynamic depth} is the former less the latter~[28].  Highly regular
systems or those in thermal equilibrium have low or zero complexity: the
former,
because both the maximal and actual disorder are small; the latter, because
both are large but equal.  Highly complex systems have large maximal but small
actual entropies (much possible, but little actual, disorder).  The negative
of actual entropy is often called {\it negentropy} and is equivalent to
{\it information}~[29].  Thus, for a fixed maximal entropy, complexity and
information are equivalent.

Care must be taken in applying the second law to the biosphere; because it is
not a closed system, its entropy can decrease.  Biological, and in
particular, economic, activity acts as a limited Maxwell demon in
decreasing local entropy as the entropy of the Universe rises~[30].
Physical processes inject or remove entropy in four ways:
(1) sunlight received; (2) heat released from the Earth's interior; (3)
radiation re-emitted into space; and (4) material and heat subducted into the
Earth's crust on the ocean floors~[31].  These four, like all manifestations of
spontaneous self-organization through feedback,
are consequences of the universal struggle of
gravity and entropy.  The second and fourth are minor components of the
Earth's heat balance; with a small but important corrections from the second,
the first and third account for almost all of the Earth's heat budget~[32].
There is one further avenue for entropy flow: life stores, rearranges and
releases entropy.  Inasmuch as the complexity of life on Earth has increased
over its three-and-a-half billion year history, information stored in genes,
metabolism, form, and, more recently, in culture, life has expelled entropy
from its domain, in addition to changing the chemistry of the atmosphere.
This highlights the fact that, over long periods, the
biosphere is not at all in equilibrium, with occasional large departures from
stasis occurring against a background of steady, smaller changes~[33,34].
For metabolism to occur, however, entropy also has to be released, as with
the consumption of food or the burning of hydrocarbon fuels.  The instantaneous
effect on the Earth's gross heat balance is small, a slight Gaian fever of
global warming, although the consequences over long times may be substantial.

In examining the entropy produced by humans, we can distinguish two broad
classes of economic activity, one involving the withdrawal, the other the
release, of entropy.  The latter class I call {\it consumables} or {\it fuels,}
such as food, hydrocarbons, etc., that are taken from a state of greater
complexity and subject to chemical reactions that release energy and entropy,
bringing them closer to thermodynamic equilibrium.  Consumption, the goal of
all economic activity, is the destruction of value and demand.
The energy released
heats or performs work.  The possibilities of recycling for this
class are quite limited, so that reduction of the associated pollution
requires greater efficiencies of production and use to minimize entropy
production.  The former class I call {\it durables.}  These include
packaging --- containers, newspapers, clothing, housing, etc. --- as well
as all manner of activities and machines that extend human powers, both for
consumption purposes --- cars, newspapers, etc. --- and capital purposes ---
tools.  Capital investment constitutes a special kind of durable, the economic
analogue of catalysis, goods that produce other goods or services yet
themselves not
consumed.  The creation of durables, like the evolution of life on the planet,
requires a large {\it increase} in complexity to create the prototypes,
which can then be copied (mass production or reproduction) with only small
additional increments of complexity (due to the copying process itself, not
to the copies~[28]).  Production is the creation of value and demand.
Because they degrade slowly, durables are recyclable,
although their use requires fuels.
The major limitation is design, embodied in the original of
the item and reflected in its mass copies.  Of course, some durables we already
recycle; we use houses, clothing, and cars over and over, rather than produce
new ones for each use.  The initial rarity of complex objects and processes
creates opportunities for profit and economic growth, which proceeds by the
exploitation of new and generally temporary relative scarcities.  The
scarcities usually disappear because competition and/or mass production (if
increasing returns appear~[22]) turns the rare into the commonplace,
eliminating the relative advantage of scarcity and its
above-normal profits.  A new cycle of growth can then begin only with the
discovery of another rarity.  Economic equilibrium is a thermodynamic
non-equilibrium steady state (steady flows of matter and energy),
while economic growth departs further from thermodynamic equilibrium into
non-steady states of matter and energy flow~[35].

The reduction of entropy production from fuel use is subject to stringent
thermodynamic limits.  The burning of fuel to produce heat automatically
produces entropy, although the heat can be trapped and preserved for a time
by insulation.  The case of transforming heat into work is more subtle.
As a simplified example, consider the canonical two-temperature heat engine.
The ideal efficiency $\eta$ of such an engine burning fuel at absolute
temperature $T_h$ and releasing heat into an environment at absolute
temperature $T_c,$ where $T_h\ge T_c,$ is:
$$\eqalign{
\eta = 1 - T_c/T_h,}\eqno{(1)}
$$\noindent
where $\eta$ is the fraction of heat that can be converted to useful work.
Note that when the two temperatures are equal, no useful work can be extracted.
The transformation of heat to work requires a temperature {\it difference.}
Furthermore, even in the ideal case, when the {\it net} entropy produced is
zero, a gross entropy increment is still added to the environment, taken from
the fuel.  For a
relatively fixed environmental temperature $T_c\simeq 286\ ^\circ$K
(13 $^\circ$C or 56 $^\circ$F), the ideal efficiency of heat engines can be
increased only by raising the burning temperature $T_h.$  Thus, coal is more
efficient than wood, petroleum than coal, natural gas than petroleum, and
nuclear fusion than nuclear fission, under ideal conditions, each process
burning at a successively higher temperature.  The greater ideal efficiency
of successive fuels is also demonstrated by a detailed study of their
reactions;
the thermodynamically more efficient burners burn further toward completion
of their respective reactions.  The final products of hydrocarbon burning are
carbon dioxide and water; the final products of nuclear burnings are nuclei
closer to the most stable nucleus, iron-56, than the reactant nuclei, uranium
and plutonium splitting to lighter nuclei, and deuterium and tritium fusing
to helium.  The less efficient reactions not only do not move as far toward
completion, they also produce unburned intermediate products that are often
harmful, such as nitrogen dioxide, carbon monoxide, and unstable (radioactive)
fission products.  Nuclear fusion, burning at a higher temperature, produces
none of the unstable isotopes that fission produces.
(Carbon dioxide itself still poses the hazard
of a greenhouse effect or asphyxiation, so it must be burned in a final
step, to carbon and oxygen, by plants.  Thus the oxygen-burners depend on the
CO$_2$-burners to recycle their waste by fixing carbon.
Natural or artificial photosynthesis
could be implemented at the source of CO$_2$ emissions; new
chain reactions could be found to burn radioactive fission products into stable
isotopes.)

Apart from burning technologies with higher ideal efficiencies, any given
technology as implemented is rarely close to its ideal efficiency, thus
producing net entropy.  The ideal functioning of a given technology is reached
when it proceeds adiabatically.  The Nirvana of adiabaticity is achieved by
smoothing in {\it time} --- burning as slowly as possible --- and by isolating
in {\it space} --- perfect insulation between regions of differing temperature.
\bigskip
\noindent{\bf Final Thoughts}
\bigskip
The concept of minimum entropy production~[33] can be linked to the concept of
a {\it sustainable economy,} a human economy with a lifetime of indefinite
length~[36].  The biosphere, through its leakages of entropy, is capable of
eliminating entropy at no more than a certain, but unknown, rate.  This
putative limit is, speaking broadly, the sustainable limit to the activity of
life and is sometimes called the {\it carrying capacity}~[36] of the
biosphere.  The biosphere can be burdened only to a certain limit before it
undergoes rapid, radical, and probably unpredictable changes~[33,37].  Apart
from the pollution of commons and the appropriation of wilderness owned by no
one, the chief source of environmental degradation today, as throughout human
history, is the discriminatory use of political power to remove the burden of
resource cost from resource users by subsidy and political force and to place
it on others,
so that these users fail to pay the cost of their activities and
consequently overexploit the resources at their disposal.  Absent such subsidy
or political force and with politically guaranteed property rights of various
types, ecological sustainability of the human economy is automatic, as human
economic activity (making the most with the least) is only a special case of
a principle universal to all organic evolution, that of minimal entropy
production~[33,35]: Nature is usually economical, in the sense of parsimony.
With accurate,
market-clearing prices for natural resources, prices are linked
directly to cost and thence to waste (entropy) production~[38].  Such an
economy does not need to be made part of Nature, because it already is, as the
humans who make it up have always been.  The great exception to this principle
of ``most with least'' among humans is represented by war and war-like
collectivist economies, such as feudalism, socialism, communism, and the
various types of fascism and statism, whose participants, under the spell of
the power principle or the desperation of war, consider benefit without
registering cost.  Perhaps the major philosophical difficulty preventing
us from seeing the truth about evolution and human economies are the
still-common fallacies of social Darwinism and sociobiology: in fact, culture
is adaptively learned, not genetically programmed, and most
evolution is peaceful competition for ecological niches, analogous in every
way to peaceful economic competition, contrary to the famous but misleading
and bloody predator-prey picture of ``Nature red in tooth and claw''~[2,39].
Indeed, human history is merely organic evolution at work in culture and
economy.

It may seem as if the development of modern surplus-trading economies itself
has led humanity out of an idyllic relation to Nature and that the solution to
our growing ecological difficulties is a return to the subsistence mode of
existence that would require a far smaller and poorer population than at
present~[40].  However, with occasional exceptions, there has never been an
innocent relation between humans and their environment, at least since the
invention of agriculture: often not knowing their own strength, they have
usually exploited it up to the limits of their available technology and the
local carrying capacity~[41].  Our present situation is a result of the
cumulative inertia of thousands of years of technological progress.  The
difference today is that, unlike previous eras, the power of technology and
human numbers has grown so great that it, like nuclear weapons, threatens the
entire global ecosystem for the first time in history.  But the same
technology that damages the biosphere can also measure and heal that damage.
Minimizing this harm requires closing the natural resource cycle of modern
economies and instituting policies to extend property rights and eliminate
subsidy that blocks resource cost from individual economic decisions --- in
short,
the application of standard mechanisms to an activity never fully rationalized
before in economic terms, the production, disposal, and reuse of waste.
\bigskip
\noindent{\bf References and Notes}
\bigskip
\item{1.} Oscar Wilde, {\it Lady Windermere's Fan} (1892) in {\it Plays, Prose
Writings, and Poems} (Everyman's Library, Alfred Knopf, New York, 1991).

\item{2.} M. Rothschild, {\it Bionomics: The Inevitability of Capitalism}
(Henry Holt \& Co., New York, 1990); new ed., {\it Bionomics: Economy as
Ecosystem} (Henry Holt \& Co., New York, 1992).  A classic study of the
entropic nature of life is: Erwin Schr\" odinger, {\it What Is Life? The
Physical Aspect of the Living Cell} (1944) (Cambridge University Press,
Cambridge, 1992).  For the bridge connecting non-equilibrium thermodynamics and
the law of compensatory change to organic evolution (and thence to
human economics; see refs.~[33-35] below), see: D. R. Brooks, E. O. Wiley,
{\it Evolution as Entropy: Towards a Unified Theory of Biology,} Science and
Its Conceptual Foundations, D. L. Hull, Ed. (University of Chicago Press,
Chicago, 1986).  A biological example of compensatory change is homeostasis
in warm-blooded animals.

\item{3.} Adam Smith, {\it An Inquiry into the Nature and Causes of the Wealth
of Nations} (1776) 2 vols., R. H. Campbell, Ed. (Oxford University Press,
Oxford, Glasgow ed., 1976).

\item{4.} Historically, the most important were the ideas of Ricardo and
Malthus; see below.

\item{5.} R. U. Ayres and I. Nair, {\it Physics Today} {\bf 37}, 62
(November 1984).

\item{6.} P. Wonnacott and R. Wonnacott, {\it Economics} (McGraw-Hill, New
York, ed. 2, 1982).

\item{7.} The most basic of human tools are the now-innate ones first evolved
in hominid history, hands and language (more generally, conceptual thought).
Apart from these, humans must produce everything they use and consume;
just as, having innate ability to learn but no innate knowledge, humans must
learn everything they know.

\item{8.} I have referred to the matter/energy flows as the ``real'' economy,
but this should not be misunderstood to mean that information and prices
are ``unreal'' in the sense of being unphysical, only in the sense that
information, unlike matter and energy, is not conserved.  All three --- matter,
energy, and information --- are physical in the sense that the presence or
absence of any one has independent causal consequences.  Matter is what
possesses mass or inertia; energy is the capacity to do work, force times
distance; information is organization or non-randomness (see below in text).
This is by way of excluding the naive scientism that regards biology and
economics as merely an affair of matter and energy flows, disregarding the
role of entropy, information, organization, and, most crucially, of
feedback effects (see refs.~[33-36] below).

\item{9.} David Ricardo, {\it Principles of Political Economy and Taxation}
(1817) (C. E. Tuttle, Boston, Everyman's Classic Library, 1911).

\item{10.} R. L. Heilbroner, {\it The Worldly Philosophers: The Lives and Times
of the Great Economic Thinkers} (Touchstone/Simon \& Schuster, New York, ed. 6,
1987).

\item{11.} Rev. Thomas R. Malthus, {\it An Essay on the Principle of
Population}
(1798, 1803) 2 vols., P. James., Ed.; {\it Principles of Political Economy}
(1820) 2 vols., J. Pullen, Ed. (Cambridge University Press, Cambridge, 1990).
The food crisis in England in the 1790s and early 1800s was greatly exacerbated
by the Napoleonic wars and the cutoff of food to England from France, and by
the perversity of the English poor relief laws.

\item{12.} John Maynard Keynes, {\it Essays in Biography} (1937) (W. W.
Norton, New York, 1963), essay on Malthus.

\item{13.} John Stuart Mill, {\it Principles of Political Economy} (1848)
(Augustus M. Kelley, New York, 1987); {\it The Subjection of Women} (1869)
(MIT Press, Cambridge, MA, 1970).

\item{14.} Henry George, {\it Progress and Poverty} (1879)
(Schalkenbach Foundation, New York, 1979).

\item{15.} Sen. Albert Gore, Jr., {\it Earth in the Balance: Ecology and the
Human Spirit} (Houghton Mifflin Co., Boston, 1992), which emphasizes
recycling.  The relative attractiveness of recycling depends on raw materials
prices, which rise and fall as part of the fifty-to-sixty year cycle
discovered by Kondratieff and extended historically by later workers.
The present decade, after the collapse of
raw materials prices in the 1980s, finds recycling in an economically
precarious situation.  On the other hand, in the 1970s, recycling
and conservation were more pressing, as raw materials prices rose rapidly.
The next such period, if the cycle holds (the previous three were 1790-1810,
1850-1870, and 1900-1920), will be in 2020-2040.  See: N. D. Kondratieff
(1926),
reprinted in {\it Readings in Business Cycle Theory} (Blakiston Co.,
Philadelphia, 1944).

\item{16.} Aristotle, {\it The Politics,} S. Everson, Ed.
(Cambridge University Press, Cambridge, 1988) book II:1-5;
D. L. Soden, {\it Tragedy of the Commons:
Twenty Years of Policy Literature, 1968-1988}
(Vance Bibliographies, Monticello, IL, 1988).

\item{17.} On the conservation of African wilderness and wild animal herds
through privatization, for example, see: R. Bonner, {\it At the Hand of Man:
Peril and Hope for Africa's Wildlife} (Random House, New York, 1993).
Most of the destruction of African wilderness and wildlife in historical times,
and particularly within the last thirty years, is due to war, quasi-militarized
economic collectivization, and their attendant famines.  Much the same can be
said of the disappearance of European wilderness and wildlife within
the last century.  A striking historical case of ecological degradation
through incessant war is found in the final three centuries of the Roman
empire, during which vast stretches of forest in the Mediterranean basin
were destroyed.

\item{18.} W. Brian Arthur, {\it Sci. Am.} {\bf 262}, 92 (February 1990).
Increasing returns to scale spread out over time, declining marginal costs, or
learning curves, justify {\it general} subsidy
only, still subject to competitive principles, but not targeted subsidies or
legal monopolies.  Furthermore, marginal costs cannot decline indefinitely in
any given case.  Increasing returns to scale is the economic form of the
thermodynamic phenomenon of autocatalysis, the repeated occurrence of which
causes non-equilibrium thermodynamic systems to step down into progressively
lower states of entropy production (see ref.~[35] below).  A familiar
example is the youth and maturation (ontogenesis) of an individual requiring
temporary support and education by its parents.

\item{19.} That is, the discriminatory use of political power
moves a society away from the win-win, non-zero-sum, profit-maximizing game
of economics towards the win-lose, zero-sum, power-maximizing game of politics
and war, the latter being zero-sum because political power is, in some sense, a
conserved quantity.  See: John von Neumann,
Otto Morgenstern, {\it The Theory of Games and Economic Behavior,} ed.~3
(Princeton University Press, Princeton, NJ, 1953);
also: Eugen von B\" ohm-Bawerk, {\it Capital and Interest, Vol.~2: Positive
Theory of Capital,} ed.~4 (1921) (Libertarian Press, Spring Mills, PA, 1959).

\item{20.} John A. Hobson, {\it Imperialism} (1902) (University of
Michigan Press, Ann Arbor, MI, 1965).

\item{21.} Lester C. Thurow, {\it The Zero-Sum Society} (Viking Penguin, New
York, 1981).

\item{22.} Alfred Marshall, {\it Principles of Economics,} ed.~8 (1920)
(Porcupine Press, Chicago, 1982).

\item{23.} It might seem that the problem with command economies is that the
central authorities form a political and economic monopoly whose power can be
abused.  However, monopoly producers have the curious property of reducing
output, of both useful product {\it and} waste, while raising both their
prices, in comparison to the situation in competition.  So the hyperpollution
of Communist economies cannot be due to monopoly power.  Rather, the
centralized command economy, lacking a price system, has a weak or non-existent
information network to make decisions and incentives to move itself toward
greater efficiency, while simultaneously concentrating enormous political
and technological power in the hands of central planners who cannot
acquire the information required to make economic decisions in a timely way.
Since prices provide information and incentive, fixed prices are
{\it mis}information and {\it mis}incentive, and the absence of prices
altogether is {\it non-}information and {\it non-}incentive.  In other words,
while the temporary failures of command economies may be due to abuse of power,
the more basic failures are caused by ignorance and laziness.

\item{24.} M. Feshbach and A. Friendly, Jr., {\it Ecocide in the USSR: The
Looming Disaster in Soviet Health and Environment} (Basic Books, New York,
1992).

\item{25.} {\it Garbage} {\bf IV}, 24 (January 1993).

\item{26.} Nicholas Georgescu-Roegen, {\it The Entropy Law and the Economic
Process} (Harvard University Press, Cambridge, MA, 1971).

\item{27.} {\it The Economy as an Evolving Complex System: Santa Fe Institute
Studies in the Sciences of Complexity, Volume V,} P. W. Anderson {\it et al.,}
Eds. (Addison-Wesley, Redwood City, CA, 1988).  For a semi-popular account,
see:  M. M. Waldrop, {\it Complexity: The Emerging Science at the Edge of
Order and Chaos} (Simon \& Schuster, New York, 1992).

\item{28.} S. Lloyd and H. R. Pagels, {\it Ann. Phys.} (New York) {\bf 188},
186 (1988).  See also: L. E. Reichl, {\it A Modern Course in Statistical
Physics} (U. Texas Press, Austin, TX, 1980).
The exact definition of thermodynamic depth is the difference of the
coarse-grained (maximal or Gibbs) entropy and the fine-grained (actual or
Boltzmann)
entropy.  The former is the logarithm of the number of microstates available
to a system with fixed macroscopic characteristics; the latter is the logarithm
of the number of microstates the system actually occupies.  In thermal
equilibrium, the equality of the two entropies stems from ergodicity, the
system spending the same time in each possible microstate; or, alternatively,
from equipartition, equal probability for each microstate.
Full thermodynamic equilibrium involves, besides thermal equilibrium (equal
temperatures), mechanical (equal pressure) and chemical (zero affinity)
equilibrium as well, complications we ignore here.
The coarse-grained entropy of the Universe (local comoving volume)
is itself increasing, although only
logarithmically, because of the cosmic expansion.  The universal fine-grained
entropy is virtually fixed, with only a tiny accumulating contribution from
energy sources such as stars and other non-equilibrium systems that are
gravitationally bound, such as our solar system.  Gravity in such systems
creates spatial gradients of temperature, pressure, and chemical potential that
make energy generation and heat flow possible.  See: Richard C.
Tolman, {\it Relativity, Thermodynamics, and Cosmology} (1934) (Dover
Publications, New York, 1987), chapter 9.

\item{29.} C. E. Shannon, W. Weaver, {\it The Mathematical Theory of
Communication} (U. Illinois Press, Urbana, IL, 1962); R. Shaw, {\it Z.
Naturforsch.} {\bf 36A}, 80 (1981).  A number of different quantities called
``entropy'', ``information'' and ``complexity'' appear in the literature
on thermodynamics, information theory, and evolution, related in meaning, but
not equivalent in definition.

\item{30.} J. Clerk Maxwell, letter to Lord Rayleigh (William Strutt), 6
December 1870, quoted in Emilio Segr\` e, {\it From Falling Bodies to Radio
Waves: Classical Physicists and Their Discoveries} (W. H. Freeman, New York,
1984), pp. 242-243; L. Szilard, {\it Zeit. Phys.} {\bf 53} (1929) 840.

\item{31.} R. A. Berner and A. C. Lasaga, {\it Sci. Am.} {\bf 260}, 74 (March
1989).

\item{32.} R. G. Fleagle, J. A. Businger, {\it An Introduction to Atmospheric
Physics} (Academic Press, New York, ed. 2, 1980).

\item{33.} I. Prigogine, {\it Nonequilibrium Statistical Mechanics}
(Wiley-Interscience, New York, 1962); I. Prigogine, {\it Introduction to the
Thermodynamics of Irreversible Processes} (Wiley-Interscience, New York, 1967);
P. Glansdorff, I. Prigogine, {\it Thermodynamic Theory of Structure, Stability,
and Fluctuations} (Wiley-Interscience, New York, 1971); G. Nicolis, I.
Prigogine, {\it Self-Organization in Non-equilibrium Systems} (John Wiley \&
Sons, New York, 1977); I. Prigogine, {\it From Being to Becoming: Time and
Complexity in the Physical Sciences} (W. H. Freeman, New York, 1980).
For semi-popular accounts, see: I. Prigogine, I. Stengers, {\it Order Out of
Chaos: Man's New Dialogue With Nature} (Bantam, New York, 1984), and G.
Nicolis,
I. Prigogine, {\it Exploring Complexity} (W. H. Freeman, New York, 1989).
A key result for our purposes is the existence and, in the
close-to-thermal-equilibrium regime, uniqueness of the steady state of
{\it minimal entropy production,} or {\it thermodynamic branch,} where the
entropy, prevented by some given constraints (gravity, for example) from
being at maximum, steadily increases toward maximum at the minimum rate
possible under those constraints.  The evolution of non-equilibrium
thermodynamic systems along the thermodynamic branch describes, without
determining in specifics, most forms of organic and non-organic growth and
evolution.  The general Walrasian equilibrium of microeconomics is identifiable
as a particular case of this steady state of minimal entropy production and
thus represents the workings of organic evolution in human economic life.  (The
Walrasian equilibrium can be extended to include negative prices.)
Farther from thermal equilibrium exist, along with the thermodynamic, other,
higher-than-minimal entropy production branches that are macroscopically
more organized, but which exhibit greater microscopic disorder; unlike the
thermodynamic branch, these branches have {\it finite} lifetimes, as they
``burn themselves up'' more quickly.  Such branches form intermediate states
between almost or exact thermal equilibrium as one extreme and purely
mechanical, zero-entropy states as the other.  Their temporary greater
macroscopic order
is more than cancelled by a positive compensatory change in microscopic
disorder; the positive excess entropy is a sign of instability.
The paradigmatic example in human
societies is war, which, in economic terms, cannot last indefinitely, as it
requires overconsumption, undersaving, or dissaving of resources (people,
property, environment) --- excess disorder.  Analogous states appear in other
species; e.g., wolves alone or in small groups versus wolves in packs.

\item{34.} S. J. Gould, N. Eldridge, {\it Paleobiology} {\bf 3}, 115 (1977);
N. Eldridge, {\it Time Frames: The Rethinking of Darwinian Evolution and the
Theory of Punctuated Equilibrium} (Simon \& Schuster, New York, 1986).
Irreversible or historical evolution occurs only because various coupled
irreversible processes occur at {\it different} rates, the differences in
such rates determining the selective forces (natural, sexual, etc.) that
govern the {\it rate} of evolution.  The {\it direction} of evolution, on
the other hand, is set by initial conditions, not by selection.  Another
pair of commonly confused concepts are {\it function,} which is historically
evolved by selection, and {\it structure,} which is accidental or
``just-so'' and generally of no evolutionary or historical significance
whatever.  Structure is largely influenced by non-selective physical forces.

\item{35.} That is, a scarcity of scarcities slows economic growth, much as
fractal diffusion requires bottlenecks as a stimulus.  For relative scarcity,
profit, and economic growth, see: Joseph A. Schumpeter, {\it The Theory of
Economic Development} (1912, 1926), R. Opie, Trans. (Harvard University Press,
Cambridge, MA, 1934), where the steady and non-steady flows are analyzed as the
circular and innovatory flows.  Given the general
argument outlined in ref.~[33] above, one can complete the thermodynamic
identification of economic equilibrium via an {\it information
theory of value} as a replacement for the defunct labor theory of value of
Locke, Petty, Smith, Mill, and Marx,
inasmuch as economic value or utility is tied to information
content or embodied organization, not merely to work expended.  Since
information = negentropy = -entropy, the branch of minimal entropy production
is, equivalently, the branch of maximal information or complexity production.
The maximal flow of information or value is then just the maximal flow of
{\it profit} or {\it net benefit} (flow of gross benefit or utility minus
flow of cost) of the Walrasian equilibrium.  The gross benefit or utility is
the gross complexity or information, while the cost is the entropy dissipated
in producing that gross benefit or utility.  In the
close-to-thermal-equilibrium regime, at least, the components of information
production have the general form: matter/energy flow times thermodynamic force.
If information is value and the matter/energy flows are identified
with the flow of goods and services, then prices are
generalized thermodynamic forces (chemical affinities or gradients of pressure,
temperature, and chemical potential).  This identification is not
surprising, inasmuch as prices usually equilibrate the flows of goods and
services to maximize net benefit.  The economy as a whole evolves because
different goods and services flow at different rates (see ref.~[34] above).
Furthermore, the process of economic
growth as described by Schumpeter involves the temporarily destabilizing
movement from one steady state to another, which takes place by the
discovery and exploitation of economic {\it autocatalysis}, capital goods
producing more capital goods.  This step transforms the thermodynamic branch
to a lower-than-before entropy production state, the sudden and temporary
appearance of excess negentropy, complexity, or
information identifiable with the temporary above-normal profits of new,
rapidly growing economic sectors.  The temporary greater macroscopic disorder
is
more than cancelled by a positive compensatory change in microscopic order;
the negative excess entropy is a sign of stability.

\item{36.} P. M. Vitousek, P. R. Ehrlich, A. H. Ehrlich, P. A. Matson,
{\it BioScience} {\bf 36}, 368 (1986); J. MacNeill, {\it Sci. Am.} {\bf 261},
154 (September 1989); W. D. Ruckelshaus, {\it ibid.,} 166; S. Schneider, P.
Boston, Eds., {\it Scientists on Gaia} (Cambridge, MA: MIT Press, 1992).
The lifetime of human economies on Earth is in principle very long, but still
limited by the lifetime of the Sun (five to ten billion years), which in turn
is set by the Sun's approach to chemical (nuclear) equilibrium.
The notion of applying thermodynamic methods to the study of human economies
is as old as thermodynamics itself and was first suggested by Sadi Carnot,
the first thermodynamicist, in discussions with his brother Hippolyte, father
Lazare, and others of Napoleon's engineers at the {\it \' Ecole Polytechnique}
of Paris (see Segr\` e, ref.~[30] above, pp. 192-201).  The use of
physical concepts in economics thereafter acquired a bad reputation, in part
because of their misuse in the crude militaristic and rationalist parody of
the engineering mentality so powerful in the ideas of Saint-Simon and Comte,
the intellectual founders of modern socialism.  Among their fallacies, which
are still in circulation, were four crucial ones: that an economy
(or an ecosystem) can be conceived of as a single, simple
machine knowable, predictable, and/or controllable by any one person or a
small group of persons, instead of as a simplified, composite abstraction
or collage; that an economy is the result of the
deliberate design and conscious will of a single or small group of persons (a
belief that might be termed {\it social creationism} or {\it
anthropomorphism});
that economic evolution is mechanistic (reversible), instead of thermodynamic
(irreversible), like all other organic evolution; and that economic response
is linear instead of non-linear, or, equivalently, that feedback effects are
unimportant.  In short, they ignored the
essential elements of time and entropy (or information).  The relation of these
pseudoscientific superstitions to centralized or command economics should be
obvious (see ref.~[23] above).  For a presentation of and attack on this
type of economic thought, see: Friedrich A. Hayek, {\it The Counter-Revolution
of Science: Studies on the Abuse of Reason} (The Free Press, Glencoe, IL,
1952).  (See also ref.~[10] above, chap.~V.)  Most nineteenth- and
twentieth-century economists have misconceived of economics as analogous to
mechanics, rather than to thermodynamics and organic evolution.  (Even many
naturalists, such as Darwin, missed the thermodynamic nature of organic
evolution; see ref.~[2] above.)  An exception was Marshall, who correctly saw
economics as a branch of biology (see ref.~[22] above).

\item{37.} An example of a carrying capacity is the infamous Laffer
curve: an economy can be taxed only to a certain point before it starts to
yield lower revenues, that point being apparently a tax rate of around
eighty percent for closed economies, with shrinkage in total output starting
at about forty percent.  (The situation in an open economy is strongly
influenced by the tax rate in other economies.)  Isolated cases of ecological
carrying capacity have been known for millenia.  The growth of world
population and production has brought to the fore the great variety and number
of distinct ecosystems, each with its own carrying capacity.

\item{38.} An application of entropy-based pricing is the case of energy taxes.
One approach is to tax different fuels weighted by their heat content, but
this approach is only half-correct, as it does not reflect the efficiency of
the fuel burning.  For {\it heating} fuels, the entropy released is
proportional to heat content, but also inversely proportional to the
environmental temperature, varying with time and location.  For fuels
producing {\it work} (electricity, transport), the entropy released varies as
the heat content, but also inversely with the {\it burning} temperature,
reflecting the greater efficiency of hotter fuels.  {\it Refrigeration}
requires a special kind of engine, to force heat to flow from cold to hot, and
the entropy released is proportional to the heat content of the fuel {\it plus}
the insulation-adjusted temperature difference between the inside and outside
of the refrigerator and inversely proportional to the outside temperature.
The entropy-based tax is a more complicated but more accurate measure of
environmental damage.  Yet another approach is to tax fuels in proportion to
their prices, so as not to change their relative utilizations.  Although the
demand for fuels is related to their entropy release, proportionality of
entropy release to price holds only in the overidealized case of fixed fuel
supplies.  Except for this qualification, price-proportional and
entropy-release-proportional energy taxes are identical.

\item{39.} The ``tooth and claw'' phrase was popularized by Herbert Spencer.
The predator-prey-vegetable food chain is unstable, because its only steady
state is the extinction of both animal species.  Most evolution involves
peaceful competition, coexistence, or co-operation of species coevolving in a
given ecosystem.  The spread in the last century of such slogans
as ``survival of the fittest'' (another Spencerism) has done incalculable
damage to the proper understanding of evolution; more accurate slogans might
be ``elimination or modification or departure to elsewhere of the least fit (to
a given ecology)'' and ``survival of the at-least-adequate'' --- more clumsy,
but much less misleading.  Exposing the errors of sociobiology would require
a separate paper all by itself (see ref.~[2] above).  Suffice
it to say that humans are {\it not} genetically programmed to {\it act} in
a fixed way, but rather are genetically programmed to {\it learn to act} in
various adaptive ways (within the constraints set by previous evolution),
often in response to unforeseen circumstances.  Genetic programming by
selection is usually too slow a process to bear the explanatory weight that
sociobiologists place on it.  The major genetic selective effect operating on
individual animals, including humans, is not {\it natural} (death) but
{\it sexual} (choice of mate).  Apart from that, selection in human societies
operates on cultural, economic, and political patterns, not on
individuals and their genes, and on much shorter timescales.

\item{40.} J. Davis, Ed., {\it Earth First! Reader: Ten Years of Radical
Environmental Journalism} (Gibbs Smith, Layton, UT, 1991).  The underlying
fallacy of radical environmentalism is the concept of {\it preservation,} as
opposed to {\it conservation,} the false assumption that the biosphere was
in idyllic stasis until disrupted by humans (implicitly and falsely viewed as
unnatural and, at the same time, having unlimited power over Nature)
and can be returned to stasis:
in fact, there never was any stasis, nor is it possible to achieve now.
Conservation, on the other hand, correctly views ecological and human economic
activity as parts of a single whole, similar to the picture outlined in this
paper (see also ref.~[2] above).
A distinct but equally fallacious and dangerous idea is that
individual economic activity as such is the basic threat to ecological
sustainability; see, for example: G. Hardin, {\it Living Within Limits:
Ecology, Economics, Population} and D. Worster, {\it Wealth of Nature}
(both: Oxford University Press, Oxford, 1993).
The proposed cure, ecological engineering by collectivized dictatorship, is
far worse than the disease, however, being subject to the fatal flaws all such
systems face (see refs.~[22,36] above).  The so-called law of unintended or
perverse consequences enters here as a special case of compensatory change:
the attempt to suppress entropy production altogether in reality forces a
system closer to thermal equilibrium (heat death) and thus increases entropy
production.  A familiar example is the runner who attempts to save time by
sprinting but who quickly exhausts himself and ends up taking more time than
he would have otherwise.

\item{41.} See, for example: B. M. Fagan, {\it The Great Journey: The Peopling
of Ancient America} (Thames \& Hudson, London, 1987), on the ancient Americans'
use of slash-and-burn forest-clearing.
\vfill\eject
\noindent{\bf Acknowledgements}
\bigskip
The author would like to thank James Gelb of Fermilab (now of Morgan Stanley)
for helpful discussions and assistance with the figures, and Stephen Selipsky
of Yale University and George Hockney of Fermilab
for important suggestions.  This article is dedicated to the memory of the late
Friedrich A. Hayek (1899-1992) of the Universities of Freiburg and Chicago.
Work supported by the U.S. Department of Energy under contract number
DOE-AC02-76-CHO-3000.
\bigskip
{\it ...there wasn't one kind of Strife alone, but two all over the Earth.  As
for the one, a man would sing her praises when he came to understand her; but
the other is nasty: and they are completely different in nature.  One incites
war and battle, being cruel: no man loves her; but perforce,
through the will of the immortal gods, men pay harsh Strife her due honor.  But
the other is the daughter of dark Night; and the son of Kronos, who sits above
and dwells in the thin air, set her in the roots of the Earth: and she is far
kinder to men.  She rouses even the shiftless to toil; for a man grows eager
to work when he considers his neighbor, a wealthy man who hastens to plow and
plant and put his house in good order; and neighbor vies with neighbor
as he hurries after riches.}\medskip
\hfill{\it Hesiod, {\rm Works and Days}}
\bigskip
\noindent{\bf Figures}
\bigskip
{\it Figure 1.}  Conventional economy.  Investment and consumption form
circular flows from wages and profits, respectively; natural resource inputs
purchased with rents, waste output discharged at no monetized cost or benefit.
Solid lines: physical flows; dashed lines: money flows.
\bigskip
{\it Figure 2.}  Pay-for-trash-removal economy.  Same as Figure~1, but waste
discharge now monetized at negative price; producers of waste pay consumers
of waste to remove it.
\bigskip
{\it Figure 3.}  Recycling economy.  Same as Figure~1, but waste is now
purchased by consumers from producers at positive price for recycling; cost is
effectively paid by natural resource rents, and the resource-waste cycle
becomes closed.
\bigskip
{\it Figure 4.}  Instantaneous
supply-demand equilibrium for waste production.  Supply curve
is marginal cost of waste, subject always to diminishing returns (rising
marginal cost); demand curve is marginal utility of waste, typically also
subject to diminishing returns (falling marginal utility).  Two scenarios:
pay-for-trash-removal economy, wherein marginal utility of waste is mostly or
completely negative, and waste sells at a negative price; recycling economy,
wherein marginal utility of waste is mostly or completely positive, and waste
sells at a positive price.  Marginal cost of waste is partially negative,
reflecting falling total costs for moderate amounts of pollution.
\vfill\eject
\end